\documentclass{elsarticle}
\usepackage{latexsym}
\usepackage{longtable}
\usepackage{float}
\usepackage[latin2]{inputenc}
\usepackage{graphicx}
\usepackage{ae}
\usepackage{aecompl}
\usepackage{setspace}
\usepackage{amsmath}
\usepackage{amssymb}
\usepackage{amsthm}
\usepackage[hidelinks]{hyperref}

\makeatletter
\def\ps@pprintTitle{%
	\let\@oddhead\@empty
	\let\@evenhead\@empty
	\def\@oddfoot{\centerline{\thepage}}%
	\let\@evenfoot\@oddfoot}
\makeatother

\begin{document}

\title{\Large Good Neighbors, Bad Neighbors: The Frequent Network Neighborhood Mapping of the Hippocampus Enlightens Several Structural Factors of the Human Intelligence on a 414-Subject Cohort}

\author[p]{Máté Fellner}
\ead{fellner@pitgroup.org}
\author[p]{Bálint Varga}
\ead{balorkany@pitgroup.org}
\author[p,u]{Vince Grolmusz\corref{cor1}}
\ead{grolmusz@pitgroup.org}
\cortext[cor1]{Corresponding author}
\address[p]{PIT Bioinformatics Group, Eötvös University, H-1117 Budapest, Hungary}
\address[u]{Uratim Ltd., H-1118 Budapest, Hungary}

\date{}

\begin{abstract}
		The human connectome has become the very frequent subject of study of brain-scientists, psychologists, and imaging experts in the last decade. With diffusion magnetic resonance imaging techniques, unified with advanced data processing algorithms, today we are able to compute braingraphs with several hundred, anatomically identified nodes and thousands of edges, corresponding to the anatomical connections of the brain. The analysis of these graphs without refined mathematical tools is hopeless. These tools need to address the high error rate of the MRI processing workflow, and need to find structural causes or at least correlations of psychological properties and cerebral connections. Until now, structural connectomics was only rarely able identifying such causes or correlations. In the present work, we study the frequent neighbor sets of the most deeply investigated brain area, the hippocampus. By applying the Frequent Network Neighborhood mapping method, we identified frequent neighbor-sets of the hippocampus, which may influence numerous psychological parameters, including intelligence-related ones. We have found neighbor sets, which have significantly higher frequency in subjects with high-scored Penn Matrix tests, and with low-scored Penn Word Memory tests.  Our study utilizes the braingraphs, computed from the imaging data of the Human Connectome Project's 414 subjects, each with 463 anatomically identified nodes.
\end{abstract}

\maketitle
	
\section*{Introduction}

Our brain contains approximately 80 billion neurons, each connected to hundreds or even thousands of other neurons. All brain functions are closely connected to this network of the brain, frequently called ``the connectome'' \cite{Seung2009,Sporns2005,Lichtman2008a}. Today, the neuronal-level connectome (or braingraph), where the nodes correspond to the 80 billion neurons, and two nodes are connected by an edge if the corresponding neurons are connected by an axon, is unknown for us. The only fully developed species with known neuronal-level braingraph is that of the nematode {\em Caenorhabditis elegans}, with 302 neurons, determined in the eighties by electron-microscopic techniques (\cite{White1986}, the graph can be downloaded from \url{braingraph.org} \cite{Kerepesi2016b}). More recently, serious developments are reported in the mapping of the neuronal-level braingraph of the fruitfly {\em Drosophila  melanogaster} with 100,000 neurons \cite{Zheng2018}.

With currently available techniques the human braingraph can be constructed and analyzed in a much coarser resolution than the neuronal level, with the help of diffusion magnetic resonance imaging (MRI) \cite{Hagmann2012}. In these graphs, the nodes are anatomically identified 1-1.5 cm$^2$ areas of the gray matter (frequently addressed as ``ROIs'', i.e., Regions Of Interests), and two nodes are connected by an edge if the diffusion MRI analyzing workflow \cite{Hagmann2012,Fischl2012,Gerhard2011,Daducci2012} finds axonal fiber tracts between them. Therefore, we can construct today braingraphs upto 1015 nodes and several thousand edges. Perhaps the most reliable large human MRI datasets to date are the public releases of the Human Connectome Project (HCP) \cite{McNab2013}. 

\subsection*{The graph-theoretical analysis of the braingraph}

The exact, robust and graph-theoretical analysis of the human braingraphs is a fast developing and important area today. Our research group has contributed numerous results in this field, analyzing the HCP data. We have computed hundreds of braingraphs \cite{Kerepesi2016b}, and prepared the Budapest Reference Connectome Server, which generates the graph of $k$-frequent edges of the human connectome of n=477 people, where $1\leq k\leq n$, and the $k$-frequent edges are those, which are present in at least $k$ braingraphs out of the n=477. The parameter $k$ is selectable, along with other parameters at the webserver \url{https://pitgroup.org/connectome/}, and the resulting consensus graph can be visualized and downloaded from the site \cite{Szalkai2015a,Szalkai2016}.

In the work \cite{Kerepesi2015a}  we have mapped the individually more and less variable lobes of the human brain on 395 subjects, with the help of a natural measure: the distribution function. We have shown that the frontal and the limbic lobes are more conservative, while the edges in the temporal and occipital lobes show more diversity between the individual braingraphs. We have also compared the lobes of the brain by computing numerous graph-theoretical parameters in the sub-graphs, induced by the vertices of the lobes in \cite{Szalkai2017c}. We have found that the right temporal and the right parietal lobes have better connectedness-related graph-theoretical parameters than the left ones (e.g., larger minimum vertex cover, larger Hoffman-bound). More interestingly, the left frontal lobe has better such parameters than the right one. 

We have compared the volumetric properties of the male and female brain areas in \cite{Szalkai2016b}, and the sex differences in the human brain connectomes in \cite{Szalkai2015,Szalkai2016a,Szalkai2015c}. We have shown a strong statistical advantage of the female connectomes in the connectedness-related advanced graph-theoretical parameters in a smaller cohort in \cite{Szalkai2015} and in a larger cohort in \cite{Szalkai2016a}. In \cite{Szalkai2015c} we have clarified that the better, connectedness-related braingraph parameter-results of women cannot be due to the brain-volume differences: we have identified 36 large-brain females and 36 small-brain males, such that the brain volumes of all females were larger in the group than those of all males, and the advantage of the women remained valid even after this highly specific subject selection. 

The development of the connections in the mammal brains is a hot research area today with many open questions. Lots of information were learned from embryonic rat and mouse brain microscopy on the development of single neuronal tracts \cite{DeCarlos1992,Nonomura2013}. In the human brain, much less is known about the phases of the axonal development and growth. By analyzing the features of the publicly available Budapest Reference Connectome Server \url{http:\\connectome.pitgroup.org}, we have discovered the phenomenon of the Consensus Connectome Dynamics (CCD), which, by our hypothesis, describes the individual axonal development of the human brain \cite{Kerepesi2016,Szalkai2016e,Kerepesi2015b,Szalkai2016d}. The CCD phenomenon is also applicable for directing the edges of the braingraph \cite{Kerepesi2015b,Szalkai2016d}.

\subsection*{Robust methods}

The robust analysis of the MR imaging data is an important point in all applications, since there are numerous complex steps, where noise or data processing artifacts may appear in the image processing workflow. For example, one such area is the tractography phase, where the crossing axonal fibers may induce errors in the processing \cite{Zhan2015,Jbabdi2011,Bastiani2012}. Therefore, the error-correcting analytical methods have an utmost importance in processing of these data.

Our research group pioneered several such methods by examining the frequently appearing substructures. This approach will not consider rarely appearing errors, since if we deal with substructures, which appear with a minimum frequency of 80\% or 90\%, then the infrequent errors will be filtered out. The Budapest Reference Connectome Server generates the $k$-frequent edges \cite{Szalkai2015a,Szalkai2016}. In the work \cite{Fellner2017} we have mapped the frequently appearing subgraphs of the human connectome. The frequent complete subgraphs of the human braingraph were identified in \cite{Fellner2019}. 

Numerous publications attempt to find correlations between the psychological and anatomical, more exactly, connectomical, or graph-theoretical properties of the braingraph (e.g., \cite{Szalkai2016c}).
The difficulty of identifying structural-psychological correlations lies in the individual diversity of the cerebral connections. One possible solution to this difficulty is the comparison of the {\em frequent} substructures with the results of psychological measurements. 

In the publication \cite{Fellner2018} we defined the Frequent Network Neighborhood Mapping.

\subsection*{The Frequent Network Neighborhood Mapping}

Here we would like to formalize the frequent neighborhood mapping. The motivation of the formalism below is the identification of the robust, frequent neighborhoods of some important node $u$, where the word ``frequent'' means that the same neighborhood of $u$ appears frequently in the braingraphs of the N subjects of ours:

Let $G(V,E)$ be a graph with vertex-set $V$ and edge-set $E$. Let $u$ be a vertex. Vertex $v$ is a neighbor of $u$ if the unordered pair $\{u,v\}$ is an edge of $G$. Then $\Gamma(u)$, called the neighbor-set of $u$,  contains all the neighbors of vertex $u$, that is:
$$\Gamma(u)=\{v\in V: \{u,v\}\in E\}.$$

Now, let us consider $N$ graphs $G_1(V,E_1),G_2(V,E_2),\ldots,G_N(V,E_N)$ on the very same vertex-set $V$. Let $u\in V$, and let 
$$\Gamma_i(u)=\{v\in V: \{u,v\}\in E_i\}, \hbox{ for } i=1,2,\ldots,N.$$

In other words, $\Gamma_i(u)$ is the neighborhood of $u$ in graph $G_i$. 

We say that the vertex-set $W\subset V$ is a $k$-frequent neighborhood of $u$ if there are at least $k$ indices $i$, such that $W\subset \Gamma_i(u)$. If, say, $k/N\geq 0.8$, then $W$ is a frequent neighbor set of $u$ with a cut-off value (or threshold) of 80\%.

In the work \cite{Fellner2018} we have identified the frequent neighbor sets of the hippocampus of size at most 4, with threshold of 80\%. We have also identified the frequent neighbor-sets of the hippocampus, which were more frequent in male and female subjects, respectively. 

\section*{Discussion and Results}

The hippocampus is, perhaps, the most frequently and deeply investigated area of the brain: it is a part of the limbic system, it has a role in turning short-time memory into long-time memory, in spatial orientation, navigation and memory \cite{Santin2000,Voineskos2015,Nees2014,Iglesias2015}. It is a sea-horse-shaped organ, and it is present in the left- and also in the right hemispheres: that is, there is a left- and a right hippocampus in the brain. 

Here we identify the frequent hippocampus neighbor sets of size up to 4, for hippocampi in both hemispheres. Next, we investigate whether the presence of these neighbors of the hippocampus has any statistical significance with some, intelligence-related test results of the subjects. We call the hippocampus neighbor-sets, with these significant differences in frequencies "significant sets" in short.

The motivation of this study is as follows: by the best of our knowledge, no connections were proven between the presence or absence of any {\em single} connectome-edge and any psychological property of the subjects examined. This failure may be due to the great variability and plasticity of the brain connections \cite{Kerepesi2015a,Szalkai2015a,Szalkai2016}. Here we want to overcome these difficulties in a two-fold strategy:

\begin{itemize}
	\item[(i)] Instead of the individual appearances of graph-theoretical objects we consider frequent objects;
	\item[(ii)] Instead of frequent single edges from vertex $u$ we consider frequent subsets of the neighbor-set $\Gamma(u)$, where $u$ is the hippocampus.
	\end{itemize}

\subsection*{Measures of intelligence}

In the present study, we consider two psychological tests, which were administered to the subjects of the Human Connectome Project:

PMAT24\_A\_CR: Penn Matrix Test: Number of Correct Responses; scored from 0 through 24. This is a multiple-choice test where the subject needs to choose the best fit from a list of objects into the one empty position of a small matrix of objects. The PMAT test is believed to assess the mental abstraction and flexibility \cite{Gur2001}. The higher scores show better mental abilities. We grouped the scores as ``low'' between 0 and 16, and ``high'' between 17 and 24; the cut-off score 17 is the median.

IWRD\_TOT: Penn Word Memory Test: Total Number of Correct Responses, scored from 0 through 40. In the first phase of the test, the subjects need to memorize 20 written words. In the recognition phase, 40 words are shown, and the participants need to decide whether the words were seen in the first phase or not. The score is the number of the correct answers. We valued the scores 0-35 as ``low'' and 36-40 as ``high'', the cut-off score 36 is the median.

Table 1 shows the results of the Frequent Network Neighborhood Mapping for these two tests. The table list the numbers of the frequent neighbor sets of the left- and the right hippocampus in the connectomes of the subjects with high- and low PMAT24 and IWRD test scores, respectively. 

In the columns, labeled by 1,2,3 and 4 the numbers of the 1,2,3 and 4-element frequent neighbor-sets are given, for the subjects with high and low test scores. The threshold for ``frequent'' sets is 80\% in the cases of both the right- and the left hippocampus, and 90\%, when their union is considered. The column with ``significant'' label contains the number of the neighborhood sets of the statistically differing (p=0.01) frequencies in the ``low'' and the ``high'' test scores (called briefly ``significant sets''). In the case of PMAT24 tests, the majority of the significant sets are related to the high test values. This may imply that these neighborhoods of the hippocampus are beneficial for the PMAT24 test results, so, these are the ``good neighbors'' of the hippocampus.

\subsection*{Good and bad neighbors of the hippocampus for the Penn Matrix test}

Here we list some neighbor sets with significant differences of frequencies in low- and high-scored PMAT24 subjects. The full lists can be downloaded from \url{http://uratim.com/hc/hc_neighbors_PMAT_IWRD_xls.zip}; we refer to the naming conventions of the files there to the "Data Availability" section below.

The the naming of the nodes below follows those listed in the CMTK nypipe GitHub repository \url{https://github.com/LTS5/cmp_nipype/blob/master/cmtklib/data/parcellation/lausanne2008/ParcellationLausanne2008.xls}.

\noindent In this test, most of the significant sets are related to the higher scores.
\medskip

\noindent The following three neighbor sets of the left hippocampus have a significantly higher frequency in low-scored PMAT24 subjects:
\medskip

(lh.bankssts\_3)(lh.fusiform\_5)(lh.inferiorparietal\_4)(lh.insula\_1)
\medskip

(lh.bankssts\_3)(lh.insula\_1)(lh.lateraloccipital\_9)(lh.lingual\_8)
\medskip

(lh.bankssts\_3)(lh.fusiform\_5)(lh.inferiorparietal\_4)(lh.lingual\_8)
\medskip

\noindent Two examples from the 2328 significant sets of the neighbors of the left hippocampus, having higher frequencies in the better-scored subjects:
\medskip

(Left-Caudate)(lh.fusiform\_7)(lh.inferiorparietal\_5)(lh.isthmuscingulate\_2)
\medskip

(Left-Accumbens-area)(Left-Pallidum)(lh.inferiorparietal\_5)(lh.isthmuscingulate\_2)

\medskip

\noindent Two examples from the 32 sets neighboring the united left- and right hippocampi, with higher frequencies in lower PMAT24 scored subjects:
\medskip

(lh.bankssts\_3)(lh.fusiform\_5)(lh.isthmuscingulate\_3)(rh.precuneus\_3)
\medskip

(Right-Thalamus-Proper)(lh.bankssts\_3)(lh.fusiform\_5)(lh.insula\_1)

\medskip

\subsection*{Good and bad neighbors of the hippocampus for the Penn Word Memory test}

\noindent In this test, most of the significant sets are related to the lower scores.

\noindent Two examples from the 41 neighbor-sets of the right hippocampus with significantly higher frequency in higher-scored Penn Word Memory test subjects.
\medskip

  (Right-Accumbens-area)(Right-Pallidum)(rh.bankssts\_3)
  \medskip
  
  (Right-Accumbens-area)(Right-Thalamus-Proper)(rh.bankssts\_3)(rh.isthmuscingulate\_2)
  
  \medskip
  
\noindent Two examples from the 963 neighbor-sets of the left hippocampus with significantly higher frequency in lower-scored Penn Word Memory test subjects.
\medskip

  (lh.inferiorparietal\_4)(lh.insula\_2)(lh.precuneus\_11)(lh.supramarginal\_1)
  \medskip
  
  (Left-Thalamus-Proper)(lh.inferiorparietal\_4)(lh.inferiorparietal\_5)(lh.supramarginal\_1)
 \medskip
   
\noindent Two examples from the 5484 neighbor-sets of the hippocampus with significantly higher frequency in lower-scored Penn Word Memory test subjects.
\medskip

  (lh.inferiorparietal\_4)(lh.inferiorparietal\_5)(rh.insula\_4)(rh.superiortemporal\_1)
  \medskip
  
  (Right-Thalamus-Proper)(lh.inferiorparietal\_5)(lh.precuneus\_11)(rh.superiortemporal\_1)
  
\medskip

\begin{table*}[t]
	\centering
	\scriptsize
		
\begin{tabular}{ | l | l | r | r | r | r | r | r | }
	\hline
	PMAT24 &  & 1 & 2 & 3 & 4 & significant & sign. for whom \\ \hline
	hippocampus left & high & 39 & 665 & 6646 & 42854 & 2331 & 2328 \\ \hline
	hippocampus left & low & 41 & 631 & 5164 & 25824 &  & 3 \\ \hline
	hippocampus right & high & 50 & 873 & 8142 & 48521 & 1788 & 1757 \\ \hline
	hippocampus right & low & 49 & 817 & 7059 & 37558 &  & 31 \\ \hline
	hippocampus & high & 62 & 1325 & 15297 & 113579 & 5345 & 5313 \\ \hline
	hippocampus & low & 54 & 1036 & 10761 & 70252 &  & 32 \\ \hline
	IWRD &  & 1 & 2 & 3 & 4 & significant & sign. for whom \\ \hline
	hippocampus left & high & 39 & 637 & 5684 & 31139 & 963 & 0 \\ \hline
	hippocampus left & low & 41 & 691 & 6675 & 41200 &  & 963 \\ \hline
	hippocampus right & high & 47 & 833 & 7663 & 43337 & 456 & 41 \\ \hline
	hippocampus right & low & 49 & 850 & 7705 & 43918 &  & 415 \\ \hline
	hippocampus & high & 55 & 1082 & 11219 & 72613 & 5484 & 0 \\ \hline
	hippocampus & low & 62 & 1307 & 15077 & 114860 &  & 5484 \\ \hline
\end{tabular}
\caption{The table list the numbers of the frequent neighbor sets of the left- and the right hippocampus in the connectomes of the subjects with high- and low PMAT24 and IWRD test scores, respectively. In the columns, labeled by 1,2,3 and 4 the numbers of the 1,2,3 and 4-element frequent neighbor-sets are given, for the subjects with high and low test scores. The threshold for ``frequent'' sets is 80\% in the cases of both the right- and the left hippocampus, and 90\%, when their union is considered. The column with ``significant'' label contains the number of the neighborhood sets of the statistically differing (p=0.01) frequencies in the ``low'' and the ``high'' test scores (called briefly ``significant sets''). In the case of PMAT24 tests, the majority of the significant sets are related to the high test values. In the case of the IWRD test, the majority of the significant sets are related to the low test values. }
\end{table*}

\section*{Materials and Methods}

The braingraphs in our work were computed from the MRI data of the Human Connectome Project's Public Data Release at \url{http://www.humanconnectome.org/documentation/S500} \cite{McNab2013}. The subjects of this study were healthy young adults, between the ages of 22 and 35 years. The braingraphs were computed by us, applying the CMTK toolkit \cite{Daducci2012}, with randomized seeding, 1 million streamlines and deterministic tractography. We have used the 463-vertex graphs for the present work.  The graphs are available freely for download at our site: \url{https://braingraph.org/cms/download-pit-group-connectomes/}. The workflow, by which the graphs were computed from the HCP data is described in details in \cite{Kerepesi2016b}.

The computation of the frequent neighbor sets of the hippocampus, which facilitated the Frequent Network Neighborhood Mapping, used an apriori-like algorithm  \cite{Assoc,Agrawal1994}, with small modifications: \url{http://adataanalyst.com/machine-learning/apriori-algorithm-python-3-0/}, similarly as in \cite{Fellner2018}. Succinctly, the apriori algorithm makes use of the following observation: If vertex-set $A$ is frequent with a cut-off value, say 80\%, then all of the subsets of $A$ has a frequency at least of the cut-off value, i.e., 80\%. Therefore, the 2-element frequent sets can be built from the 1-element frequent sets, the 3-element frequent sets from the already identified 2-element frequent sets, and so on.

For the identification of the frequent neighbor-sets, we have applied a two-step strategy: First, we partitioned the braingraphs into two groups by the parity of the second digit of the ID of the subjects. Next, we identified the frequent neighbor sets of the hippocampus within both classes of the partition (using the apriori algorithm), with a cut-off value of 80\%. Only those sets were accepted to be frequent, which were frequent with cut-off value 80\% in both classes. In a certain sense, this strategy modeled the frequency counting in random subsets; consequently, those neighbor sets, which were frequent only in one of the classes, were identified as such.

\subsection*{Statistical Analysis}

Next, we analyzed the appearance of the frequent hippocampus neighbor sets in the high- and low-scored PMAT24\_A\_CR and IWRD\_TOT subjects. We identified the neighbor-sets, which were significantly more frequent in the connectomes of the high-scored and the low-scored subjects. For the statistical analysis we used $\chi^2$ test with significance bound of $p=0.01$, with Holm-Bonferroni corrections \cite{Holm1979}.

Our null hypothesis is that the frequencies are the same
in the connectomes of the low- and high scored subjects, and we refute this hypothesis with p=0.01.

For a neighbor set $F$, its occurrences were counted in the low-scored dataset by $count_1(F)$ and in the high-scored dataset by $count_2(F)$.
The support was computed as follows: $supp_i(F) = \frac{count_i(F)}{S_i}$, where $S_i$, for $i=1,2$, is the number of the subjects with low- and high scores, respectively.

For the significance analysis in the difference of $supp_1(F)$ and $supp_2(F)$ we used the $\chi^2$-test for categorical data:

\begin{center}
	\small
	\begin{tabular}{ l | c | c | c }
		& contains $F$ & does not contain $F$ & total \\ 
		\hline
		1$^{st}$ sample & $count_1(F)$ & $S_1 - count_1(F)$ & $S_1$ \\  
		2$^{nd}$ sample & $count_2(F)$ & $S_2 - count_2(F)$ & $S_2$ \\  
		\hline
		total & $count_1(F) + count_2(F)$ & $S_1+S_2-count_1(F)-count_2(F)$ & $S_1 + S_2$   
	\end{tabular}
\end{center}

Now: 
\[ \chi^2 = \frac{(count_1(F) \cdot (S_2-count_2(F))-count_2(F) \cdot (S_1-count_1(F)))^2 \cdot (S_1+S_2)}{S_1 \cdot S_2 \cdot (count_1(F)+count_2(F)) \cdot(S_1+S_2-count_1(F)-count_2(F))}\]

The degree of freedom for this test is one (since it is the number of samples minus one times the number of categories minus one).

\noindent {\bf Holm-Bonferroni correction \cite{Holm1979}:} The p-values for the frequent sets were ordered $p_1 \leq p_2 \leq p_3 \leq \ldots \leq p_m$.
For a significance level $\alpha = 0.01$, let the Holm-Bonferroni value for $k^{th}$ frequent set be defined as $p_k^{'} = \frac{\alpha}{m+1-k}$.
Then let $t$ be the minimum index such that $p_t > p_t^{'}$: The null hypotheses for indices $i\leq t$ need to be rejected.

\section*{Conclusions} By the application of Frequent Network Neighborhood Mapping, we examined the neighbors of the human hippocampus, and found that some frequent neighbor sets correlate with the better Penn Matrix test results, and some frequent neighbor sets correlate with worse Penn Word Memory test results. By our knowledge, this is the first result which connects the intelligence-related measures with the neighbors of the human hippocampus, with strict statistical significance analysis.
		
\section*{Data availability} The data source of this study is Human Connectome Project's Public Data Release at \url{http://www.humanconnectome.org/documentation/S500} \cite{McNab2013}. 

The parcellation data, containing the ROI labels, is listed in the CMTK nypipe GitHub repository \url{https://github.com/LTS5/cmp_nipype/blob/master/cmtklib/data/parcellation/lausanne2008/ParcellationLausanne2008.xls}.

The braingraphs can be downloaded from the \url{https://braingraph.org/cms/download-pit-group-connectomes/} site, by choosing the ``Full set, 413 brains, 1 million streamlines'' option. In the present study, we have used exclusively the 463-node resolution graphs.

The result tables, with the listing of the frequent neighbor sets, whose frequency differ significantly in low- and high scored subjects, can be downloaded in Excel format from the site \url{http://uratim.com/hc/hc_neighbors_PMAT_IWRD_xls.zip}. The archive contains 12 files. Six of them has the prefix IWRD\_TOT, six of them PMAT24\_A\_CR, containing the significant sets for these tests. After the prefix, the filenames carry strings hc\_l or hc\_r or hc, meaning that the neighbor-sets are those of the left- or right hippocampus, or the union of those. Next, the word "lower" or "upper" mean that the neighbor sets have significant differences in the frequency in the lower half or the upper half of the scored subjects. Those tables, which correspond to 0s in Table 1 are empty.
	
\section*{Acknowledgments}
Data were provided in part by the Human Connectome Project, WU-Minn Consortium (Principal Investigators: David Van Essen and Kamil Ugurbil; 1U54MH091657) funded by the 16 NIH Institutes and Centers that support the NIH Blueprint for Neuroscience Research; and by the McDonnell Center for Systems Neuroscience at Washington University. VG and BV were partially supported by the VEKOP-2.3.2-16-2017-00014 program, supported by the European Union and the State of Hungary, co-financed by the European Regional Development Fund, VG and MF by the NKFI-126472 and NKFI-127909
 grants of the National Research, Development and Innovation Office of Hungary. BV and MF was supported in part by the EFOP-3.6.3-VEKOP-16-2017-00002 grant, supported by the European Union, co-financed by the European Social Fund.



\end{document}